\documentclass[aps,pre,twocolumn,groupedaddress,showpacs,floatfix]{revtex4}
\usepackage{amsmath}
\usepackage{amssymb}
\usepackage{graphicx}

\begin{document}

\title{Hub-Based Community Finding}

\author{Luciano da Fontoura Costa} 
\affiliation{Institute of Physics of S\~ao Carlos. 
University of S\~ ao Paulo, S\~{a}o Carlos,
SP, PO Box 369, 13560-970,
phone +55 162 73 9858,FAX +55 162 71
3616, Brazil, luciano@if.sc.usp.br}

\date{25th April 2004}

\begin{abstract}   

This article presents a hub-based approach to community finding in
complex networks.  After identifying the network nodes with highest
degree (the so-called hubs), the network is flooded with wavefronts
of labels emanating from the hubs, accounting for the identification
of the involved communities.  The simplicity and potential of this
method, which is presented for direct/undirected and
weighted/unweighted networks, is illustrated with respect to the
Zachary karate club data, image segmentation, and concept association.
Attention is also given to the identification of the boundaries
between communities.

\end{abstract}


\maketitle

The problem of community finding in complex networks
\cite{Albert_Barab:2002, Newman:2003, Dorog_Mendes:2002} represents
one of the most challenging and promising perspectives from which to
approach, characterize and understand those general structures.
Related to established areas in graph theory (e.g. \cite{West:2001})
and pattern recognition (e.g. \cite{Duda_Hart:2001, CostaCesar:2001}),
the interest in community finding in complex networks was fostered by
sociological studies (e.g. \cite{Scott:2000}) and further enhanced by
the seminal articles by Wu and Huberman \cite{Wu_Huberman:2003} and
Newman and Girvan \cite{Newman_Girvan:2004}.  The latter defined the
problem of community finding as `the division of network nodes into
groups within which the network connections are dense, but between
which are sparse'.  That work also proposed a divisive methodology
based on the concept of shortest path betweeness which has become the
main reference for community finding investigations given its good
performance, despite its relatively high computational demand.  Other
approaches include the method based on an analogy with electrical
circuits \cite{Wu_Huberman:2003}, consideration of triangular loops in
the network \cite{Radicchi_etal:2004}, application of
super-paramagnetic clustering \cite{Reichardt_Bornholdt:2004},
analysis of the spectral properties of the networks
\cite{Capocci_Caldarelli:2004}, and spectral properties of the
Laplacian matrix combined with clustering techniques
\cite{Donetti_Munoz:2004}. Despite the growing attention focused on
this issue --- see \cite{Newman_Girvan:2004, Capocci_Caldarelli:2004,
Donetti_Munoz:2004, Radicchi_etal:2004} for a good characterization of
the problem and extensive related references --- some important points
remain not completely solved, including the definition of a community
and the high computational demand implied by the most effective
techniques.

The current work describes an alternative approach to community
finding which is based on one of the most characteristic concepts
underlying the new area of complex networks, namely that of a
\emph{hub}, i.e. a node in a network exhibiting high degree.  As
emphasized by the several investigations targeting complex networks,
such nodes play determinant role in defining the connectivity patterns
in several natural structures and systems \cite{Albert_Barab:2002}.
Therefore, the consideration of hubs as starting points for network
partition represents a particularly promising perspective from
which to approach the community finding problem, a possibility which
was preliminary considered in \cite{Costa_Image:2004}.  The current
article reports on a simple and powerful hub-based community finding
methodology which involves the flooding of the network with labels
emanating with constant speed from the respective hubs.  Such a
procedure, which is related to the concept of distance transform
\cite{CostaCesar:2001, Costa_Muti:2003} in graphs
\cite{LucVincent:1990} and label propagation in orthogonal lattices
\cite{Sethian:1999, Costa_Muti:2003, CostaCesar:2001}, provides a
simple and natural means for partitioning networks, especially those
organized around hubs, into coherent communities.  Such a methodology,
as well as a post-processing step allowing integration of border
elements, is presented and illustrated in the following with respect
to three representative weighted/unweighted and directed/undirected
networks, namely the well-known Zachary karate club, image
segmentation, and concept associations.

The network under analysis is assumed to have $N$ nodes, labeled as
$i=1, 2, \ldots, N$, and $n$ edges represented as $(i,j)$, which can
have unit or general weight $w_{i,j}$ represented as $w(j,i)$ in the
respective weight matrix.  The \emph{outdegree} $O_i$ of a specific
node $i$ is herein defined as the sum of the weights of the emerging
edges, i.e. $O_i = \sum_{k=1}^N w(i,k)$, while the \emph{indegree} is
defined as $I_i = \sum_{k=1}^N w(k,i)$.  Observe that undirected
networks are characterized by $O_i=I_i$ for any $i$.  The hubs are
henceforth understood as the set of $M$ nodes with the highest
degrees.  The $d-$ball with radius $d$ centered at node $i$ is defined
as the subgraph containing all nodes which are connected to $i$
through shortest paths no longer than $d$.  The label of a specific
node $i$ can be propagated through the network by identifying the
$d-$balls centered at $i$ with subsequent distance values $d$.  If
such wavefronts are started at each of the $M$ hubs, the respective
labels are propagated as long as the nodes being reached by the
wavefronts are empty, i.e. have not been visited by another front.  In
this work, such a label propagation is performed so that the labels
emanating from the hubs with higher degree are propagated first, for
the same value of $d$, than those with lower degree.

\begin{figure}
 \begin{center} 
   \makebox[4cm][c] {
   \includegraphics[scale=0.9,angle=-90]{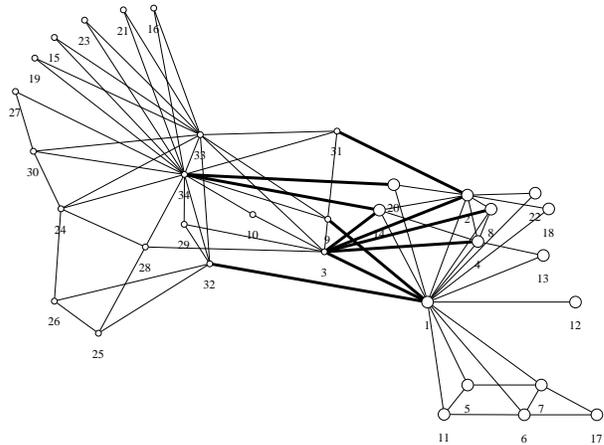} }

   \caption{Hub-based partition of the Zachary karate club network:
   the obtained communities are identified by the node size.  Only
   node 3 was labeled differently than in the original classes. The
   edge-boundary between the two communities is identified by thicker
   edges.~\label{fig:Zach}}
\end{center}
\end{figure}

The result of such a flooding procedure is to partition the original
network into $M$ communities, which can also be understood as the
Voronoi tessellation of the original network \cite{CostaCesar:2001,
Costa_Muti:2003, LucVincent:1990}.  Observe that the above procedure
implies that those nodes that are at the same distances from two hubs
are dominated by the hub with the higher degree.  Such a procedure
implies that two (or more) hubs $a$ and $b$ with $O_a>O_b$, sharing
most connections, as is the case with nodes 33 and 34 in the Zachary
club network (see Figure~\ref{fig:Zach}), may produce different
communities.  In case it is desired to merge such hubs, which is an
application-dependent decision, the following post-processing can be
performed.  For each node $i$, identify all its emanating direct
connections, whose number is represented as $E_i$, and identify the
moda (i.e. the most frequent value) $m$ among the labels of the nodes
connected to $i$.  In case the ratio $R_i$ given in
Equation~\ref{eq:R} is larger than a pre-specified threshold value
$T$, the node $i$ receives the label $m$.  For weighted networks, it
is also possible to consider the ratio between the sum of weights of
the connected nodes with label equal to the moda value and the total
sum of emerging edge weights (see Equation~\ref{eq:Rw}).

\begin{eqnarray}
  R_i = M_i / E_i \label{eq:R} \\
  R_w = \sum_{k \in moda} w(k,i) / \sum_{k=1}^{E_i} w(k,i) \label{eq:Rw} \\
  Q = \sum_i (e_{ii} - a_i^2) \label{eq:Q}
\end{eqnarray}

A particularly interesting, and somewhat overlooked, feature of a
community partition of a complex network is the \emph{boundaries}
between the identified communities.  The boundaries can be defined
with respect to nodes or edges.  In the former case, the boundary of
community $i$ can be easily identified by looking for each node with
label $i$ which is linked to at least another node with different
label.  Such a boundary, which is respective to community $i$, is
henceforth called \emph{node-boundary} of $i$.  The
\emph{edge-boundary} between two communities $i$ and $j$ corresponds
to those edges connecting nodes of $i$ to nodes of $j$ (the edge
direction can be or not observed in the case of directed networks).

Although the above described hub-based methodology can be immediately
applied to unweighted (i.e. weights are 0 or 1) or weighted networks,
some remarks regarding computational implementation should be
considered.  For undirected networks, it is more effective to follow
the subsequent connections defined by the label flooding by looking
for non-zero entries along the columns of the weight matrix and using
lists for book-keeping.  It can be verified that such a processing can
be performed in $O(N)$, as the nodes are checked only once during the
labeling procedure.  A possible means to processing weighted networks
is to visit each node while identifying the shortest path
\cite{Cormen_etal:2001} to each of the $M$ hubs, taking as result the
label of the shortest hub.  In case two (or more) hubs are found at
the same shortest path distance, that with the highest node degree is
selected.  The computational cost of finding the shortest paths
between each of the $N$ nodes and the $M$ hubs can be optimized by
using algorithms such as Dijkstra's, which implies $O(NlogN + n)$
\cite{Cormen_etal:2001}.  Effective algorithms for distance
transformation in graphs \cite{LucVincent:1990} can also considered
for further enhancing the performance.

The potential of the above described hub-based community finding
approach is illustrated in the following with respect to complex
networks obtained for the Zachary karate club, image segmentation, and
concept association.  In order to rate the quality of the obtained
communities, we consider the modularity index $Q$
\cite{Newman_Girvan:2004}.  Let the number of nodes and edges
completely contained inside community $i$ be denoted by $N_i$ and
$n_i$, respectively, and the number of edges with at least one vertex
connected to $i$ be represented as $A_i$.  The modularity index $Q$
can now be defined by Equation~\ref{eq:Q}, where $e_{ii}=n_i/n$ and
$a_i=(2n_i+A_i)/(2n)$.  Observe that $Q \leq 1$, reaching null value
for a random partition of the network \cite{Newman_Girvan:2004}.

We consider the Zachary karate club data first. The network obtained
from this dataset is often considered as a benchmark for community
finding methodologies \cite{Newman_Girvan:2004, Donetti_Munoz:2004}.
Observe that this network is unweighted (i.e. unit weights) and
undirected.  Figure~\ref{fig:Zach} shows the communities obtained by
the hub-based algorithm (small and large nodes) considering $M=2$,
followed by the above described node merging post-processing
considering $T=0.4$.  The edge-boundary between the two communities is
identified by thicker edges.  Actually, the only node misclassified by
the methodology (node 3), lies at the boundary between the two
communities and present the same number of links with both of them.
The quality of such a partition, which is precisely the same as that
obtained in \cite{Newman_Girvan:2004}, is characterized by $Q=0.36$.

\begin{figure}
 \begin{center} 
   \includegraphics[scale=0.8,angle=-90]{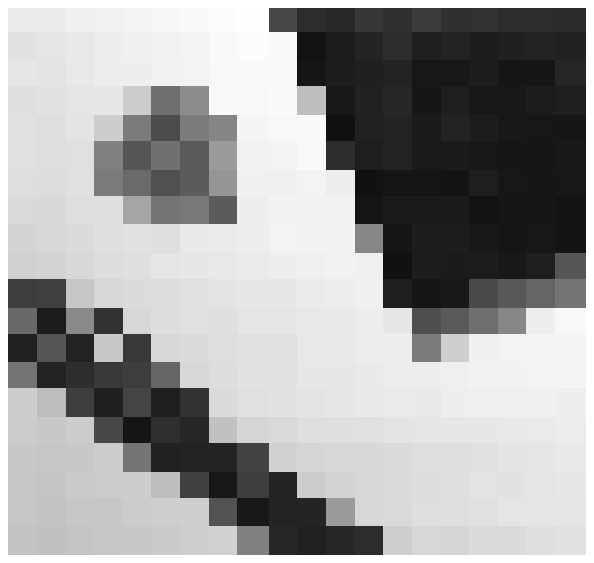}  \\
   (a) \\
   \vspace{0.2cm}
   \includegraphics[scale=0.8,angle=-90]{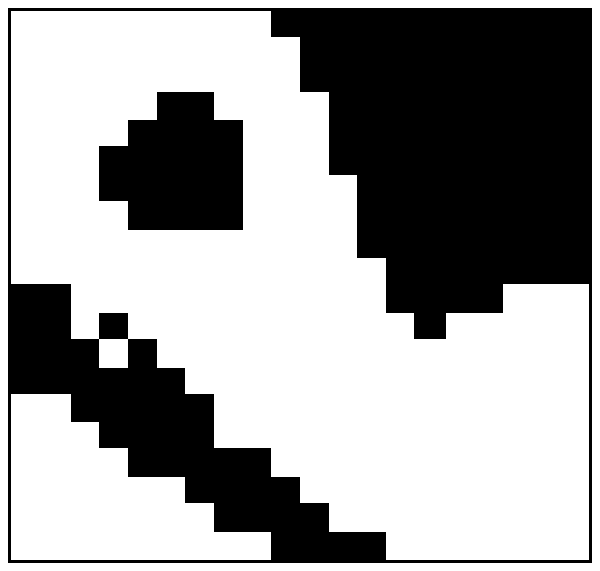}  \\
   (b) \\
   \vspace{0.2cm}
   \includegraphics[scale=0.8,angle=-90]{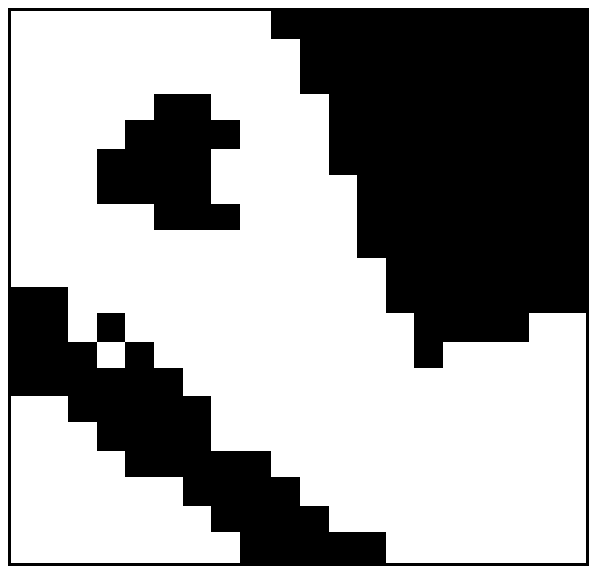}  \\
   (c) \\

   \caption{Original image (a), reference template obtained by manual
   segmentation (b), and respective hub-based segmentation into two
   regions of interest obtained by adopting $M=2$ (c).~\label{fig:img}}
\end{center}
\end{figure}

Now we draw attention to the simple image in Figure~\ref{fig:img}(a),
which contains a floppy-disk, a coin and a pencil.  The objective here
is to segment the image into reasonable regions of interest, namely
the three objects \cite{CostaCesar:2001}.  As in
\cite{Costa_Image:2004}, each image pixel is understood as a node, and
the absolute difference between the gray-levels at any two pixels
$i$ and $j$ is taken as the respective weight $w(i,j)$.  Therefore,
two pixels with similar gray-level are connected by an edge with small
weight, which can be understood as the \emph{similarity} between those
pixels \cite{Falcao:2000}. Unlike in \cite{Costa_Image:2004}, such a
fully connected graph is \emph{not} thresholded, therefore avoiding
one adjustable parameter, and the identification of the hubs is
\emph{not} performed sequentially along the processing, but as its
first step.  As such, the obtained network is weighted and undirected
(the difference between pixels is symmetric).  It should be observed
that the consideration of image segmentation as a community finding
benchmark is particularly interesting, not only because of the easy
visualization of the obtained results therefore afforded, but also for
the possibility to immediately check the coherence and quality of the
obtained communities, which should correspond to the main regions in
the original image.  In order to quantify the quality of the obtained
partitions, the template image in Figure~\ref{fig:img} is considered
as the reference for the correct classes.  Such a template was
obtained by a human operator by considering the original, higher
resolution, image from which the image in (a) was derived by
subsampling \cite{CostaCesar:2001}.  The results obtained by the
hub-based approach considering $M=2$, shown in
Figure~\ref{fig:img}(c), can be found to be in good agreement with the
template in (b).  It should be observed that, as several hubs are
obtained for the same region as a consequence of the weight-assignment
procedure (which produces a fully-connected graph as a result), the
two hubs were sampled manually from each of the two regions.  The
obtained value of $Q$ for such partitioning was found to be equal to
0.007, which is so low because of the several original connections
between the two classes implied by the procedure adopted in order to
obtain the weight matrix, which is fully connected.

\begin{figure}
 \begin{center} 
   \includegraphics[scale=0.8,angle=-90]{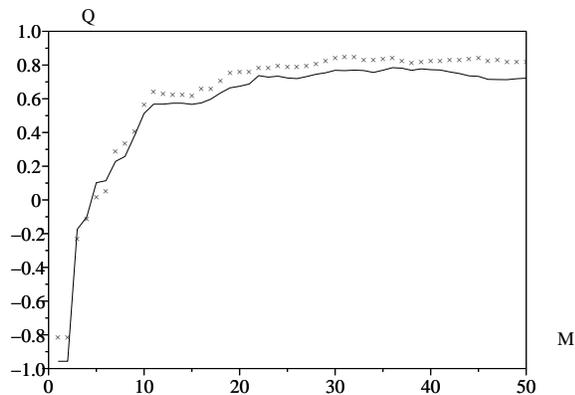}  \\

   \caption{The values of the modularity $Q$ for $M=2$ to 50 without
   (solid line) and with node-merging post-processing with $T=0.4$
   (dotted line).~\label{fig:QM}}
\end{center}
\end{figure}

Finally, we consider the concept association experiment reported in
\cite{Costa_what:2003} (see also \cite{Capocci_Caldarelli:2004}),
which involved word associations by a human subject.  A weighted,
directed network is obtained by considering each distinct word as a
node, while the weight of the edge between node $i$ and a node $j$
corresponds to the number of times the word associated to $i$ was
followed by that associated to $j$.  The hub-based community finding
algorithm was applied with $M=10$ and $T=0.4$.  Table~\ref{tab:comm}
shows five of the principal hub-words and some of the words falling on
the respectively defined communities, which include directly (shown in
italics) and indirectly associated words.  The word `fast', for
instance, was included into the community dominated by the hub {\bf
animal} through the following stream of associations {\bf animal}
$\mapsto$ \emph{butterfly} $\mapsto$ wing $\mapsto$ airplane $\mapsto$
fast.  The values of $Q$ for $M=2$ to 50 with and without the
node-merging scheme is shown in Figure~\ref{fig:QM}.  It is clear from
this curve that such post-processing is highly effective in increasing
the quality of the network partitioning.

\begin{table}
\vspace{0.7cm}
\begin{tabular}{||c|c|c|c|c||} \hline
  {\bf sun} (18) &  {\bf drink} (15) &  {\bf cold} (15) & {\bf way} (15) &  {\bf animal} (15) \\  \hline
  \emph{pyramid} & \emph{soft}& \emph{water}& \emph{easy}& \emph{cat}       \\ 
  \emph{round}   & \emph{eat} & \emph{sky} & \emph{rough}&  \emph{horse}     \\ 
  \emph{yellow}  & \emph{well}& \emph{wool}& \emph{good} &  \emph{brown}     \\ 
  \emph{circle}  & much       & \emph{air} & \emph{one}  &  \emph{butterfly}  \\ 
  \emph{hot}     & few        &  pullover  & \emph{brief}&  wing      \\ 
  triangle       &            &  sheep     & \emph{fine} &  airplane  \\ 
  drawing        &            &  thin      & \emph{single} &  fast      \\ 
   \hline
\end{tabular}
\caption{Five of the hubs with highest degree (indicated within
parenthesis) and some of the related concepts included in the
respective communities. Directly associated concetps are shown in
italics.~\label{tab:comm}}
\end{table}

All in all, the prospects of using the network hubs as references for
finding communities along the network, which can be obtained through
label propagation, has been found to provide a natural and powerful
means for partitioning complex networks, especially those organized
around hubs (e.g. scale-free networks) into coherent subgraphs.  The
potential of the reported approach has been fully substantiated with
respect to three case-examples of weighted/unweighted and
directed/undirected networks.  Given its low computational demand
(order $N$), this methodology presents good potential for several
applications in complex network research.  Future works may target
further validation of the methodology and the consideration of other
propagating schemes, such as starting the label flooding from the
nodes with the lowest degree (the end-vertices, with unit degree).  It
would be possible to assign communities to the groups of nodes
furthest away from the end-vertices, and compare such communities with
those induced by the hubs. In addition, other special nodes or
subgraphs can be considered as starting points for the flooding,
including specific paths and cycles. The latter possibility is
particularly promising for analysing networks grown around basic
cycles, such as the metabolic networks.  A particularly interesting
perspective is to explore the use of properties of the obtained node-
and edge-boundaries, such as the number of edges/nodes respectively
involved, in order to quantify the quality of the obtained results.
For instance, a small border between two regions with similar number
of nodes can be taken as an indication of high-quality community
finding. Another issue to be pursued further is to identify which
community finding algorithms are more suitable with respect to the
several types of complex networks.

\begin{acknowledgments}
The author is grateful to FAPESP (process 99/12765-2), CNPq
(308231/03-1) and the Human Frontier Science Program for financial
support.
\end{acknowledgments}

\bibliography{hubcomm}

\end{document}